# Nanoscale self-templating for oxide epitaxy with large symmetry mismatch

*Xiang Gao, Shinbuhm Lee, John Nichols, Tricia L. Meyer, Thomas Z. Ward,*

*Matthew F. Chisholm, and Ho Nyung Lee\**

*Materials Science and Technology Division, Oak Ridge National Laboratory, Oak Ridge, TN 37831, USA*

KEYWORDS

Oxide epitaxy, interfaces, VO$_2$, STEM

ABSTRACT

Direct observations using scanning transmission electron microscopy unveil an intriguing interfacial bi-layer that enables epitaxial growth of a strain-free, monoclinic, bronze-phase VO$_2$(B) thin film on a perovskite SrTiO$_3$ (STO) substrate. We observe an ultrathin (2-3 unit cells) interlayer best described as highly strained VO$_2$(B) nanodomains combined with an extra (Ti,V)O$_2$ layer on the TiO$_2$ terminated STO (001) surface. By forming a fully coherent interface with the STO substrate and a semi-coherent interface with the strain-free epitaxial VO$_2$(B) film above, the interfacial bi-layer enables the epitaxial connection of the two materials despite their large symmetry and lattice mismatch.

Epitaxial synthesis of complex oxides has stimulated considerable interest in creating novel functionalities and physical properties, through various means to control the close interactions among the order parameters, such as lattice, spin, charge, and orbital.[1-4] Heterostructures of oxide materials have also played an important role in discovering novel phenomena as they can produce well-defined interfaces to couple electronic and magnetic ground states, structure, lattice, crystallographic symmetry, etc. Most studies on the epitaxial growth of complex oxides have focused on isostructural materials, e.g. perovskites on perovskites. While for many binary oxides, such as TiO$_2$ and VO$_2$, also offer intriguing physical properties,[5-11] there are only few substrates available with similar structures (lattice parameters and crystal symmetry). The fundamental insights into the epitaxial growth of binary oxides thin films on lattice and symmetry mismatched substrates are of vital importance for exploring their unprecedented potential.[12-14]

Recently, high quality VO$_2$ polymorphs were successfully stabilized as epitaxial thin films using pulsed laser epitaxy (PLE) on perovskite substrates, such as SrTiO$_3$.[15-17] Among VO$_2$ polymorphs,[17] bronze-phase VO$_2$(B) has a monoclinic structure (with C2/m symmetry) whose lattice constants are $a = 12.03$, $b = 3.69$, $c = 6.42$ Å, and $\beta = 106.6°$,[18] whereas SrTiO$_3$ (with Pm3m symmetry) has a cubic structure with the lattice constant of 3.905 Å. Note that while many previous studies focused on R and M1 phase VO$_2$, recent studies in developing advanced energy storage found VO$_2$(B) to be a promising cathode material for Li ion batteries.[19-21] It is rather surprising that VO$_2$(B) films with corner- and edge-sharing oxygen octahedra (see **Figures 1**a and b) can be epitaxially grown on STO with corner-sharing octahedra, despite the different oxygen networks and the large biaxial lattice mismatch.

In this work, we report how two very dissimilar materials can form an epitaxial heterostructure by aberration-corrected scanning transmission electron microscopy (STEM) imaging and electron energy-loss spectroscopy (EELS). We found an interfacial bi-layer at the VO$_2$(B)/STO interface that enables epitaxial growth of a structurally complex, low symmetry film on a high symmetry substrate.

**Results and discussion**

High quality VO$_2$(B) epitaxial films were grown on (001)-oriented STO by PLE under well-optimized growth conditions. The details on the epitaxial growth and crystal quality as well as associated physical properties can be found elsewhere.[17] Figure 1 shows atomic structure projections and corresponding cross-sectional high-angle annular dark-field (HAADF) images taken along the [100]$_{VO2(B)}$ and [010]$_{VO2(B)}$ directions of VO$_2$(B). In the *Z*-contrast HAADF images, the cation columns containing Ti ($Z$ = 22), V ($Z$ = 23), and Sr ($Z$ = 38) are seen with intensities strongly dependent on their atomic number, while columns containing only light O ($Z$ = 8) atoms are hardly visible. The image shown in Figure 1d provides the reason why VO$_2$(B) is of particular interest for energy storage as the atomic structure seen from the [010]$_{VO2(B)}$ direction features an open framework that offers a good ionic diffusion pathway. The structural projection along [010]$_{VO2(B)}$ also reveals the clear symmetry mismatch between the film and substrate. Thus, we chose this orientation for the majority of the STEM investigations.

**Figures 2**a and 2b show cross-sectional HAADF images of an epitaxial VO$_2$(B) film grown on a STO substrate. The images were taken along the [100]$_{STO}$ direction. As shown in Figures 2a and 2b, the film is found to contain at least two domains aligned parallel to the [100]$_{VO2(B)}$ and [010]$_{VO2(B)}$ directions, i.e. orthogonally positioned with respect to the [100] direction of the STO substrate. In fact, the film contains two additional domains that are rotated 180 degrees about the surface normal from those imaged in Figure 2. A thin (typically ~2 nm thick) interfacial layer (IL) can be seen between the VO$_2$(B) film and the STO substrate. Based on fast-Fourier transformation (FFT) analysis, an array of misfit dislocations has formed between the IL and the structurally relaxed VO$_2$(B) film (as indicated in Figures 2a and 2b, and in the corresponding FFT images in Figure S1 in Supporting Information). The interface between the STO substrate and the IL appears to be fully coherent. As shown in Figures 2a and 2b, the average spacing between dislocations observed along the [100]$_{VO2(B)}$ direction is 3.6 ± 0.9 nm, while it is 7.9 ± 1.1 nm when seen along the [010]$_{VO2(B)}$ direction. These spacings are in good agreement with calculated values of ~3.5 nm and ~7.2 nm obtained using the lattice mismatch of +5.5 % for the [100]VO$_2$(B) ∥ [100]STO projection and -2.7 % for the orthogonal [010]VO$_2$(B) ∥ [100]STO projection. This result reveals unambiguously that the large bi-axial lattice mismatch between the film and substrate is accommodated by the creation of dislocations at the VO$_2$(B)/TL interface, i.e. strain-free VO$_2$(B) epitaxial films are obtained.

Figure 2c shows a low-angle annular dark-field (LAADF) image taken from the sample seen along the [010]$_{VO2(B)}$/[100]$_{STO}$ direction. The LAADF image highlights the interlayer, which is substantially brighter than the film or the substrate. This brighter contrast in a LAADF image indicates that the IL has a higher level of structural disorder, which leads to the electron dechannelling of the incident beam.[22-24] Based on a geometric phase analysis (GPA), it is also seen that the VO$_2$ in the IL undergoes a significant lattice expansion along the film surface normal as compared to the VO$_2$(B) film (see Figure S2 in Supporting Information). This result is consistent with the rather large in-plane compression (-5.5%) of the VO$_2$(B) film in the [100]$_{VO2(B)}$/[100]$_{STO}$ projection that will cause the observed out-of-plane expansion.

**Figure 3**a shows a HAADF image of the IL taken along the [010]$_{VO2(B)}$ direction. While there is a region of the IL (Figure 3b) that clearly duplicates the projected structure of the relaxed VO$_2$(B) film above it (except that it is rotated 180 degrees about the surface normal), most of this layer (and its FFT, Figures 3d,e) looks to be a superposition of [100], [-100], [010], and [0-10] projections of epitaxially strained VO$_2$(B). The other important feature of the IL is the extra atomic layer between the STO and VO$_2$(B) indicated with a black arrow in Figures 3b and 3c. The atomic layer shows a periodic, but different arrangement of B-site atoms than that of the TiO$_2$-terminated STO surface. The HAADF images show the out-of-plane lattice spacing between the topmost TiO$_2$ layer of STO and the extra (Ti,V)-O layer to be 2.4 ± 0.1 Å, which is significantly larger than the 1.9Å (001) plane spacing in STO. The intensity variations indicate that, in this [100]$_{STO}$ projection, the extra layer contains additional (Ti,V) columns with roughly 1/2 the B-site density of its neighboring atom columns.

Spatially resolved STEM-EELS data from the interfacial region is presented in **Figure 4**. Figures 4a and 4b respectively show element maps using the Ti-$L_{2,3}$ and V-$L_{2,3}$ signals taken from the same interfacial region shown in Figure 4c. The V-$L_{2,3}$ signal in Figure 4a shows a chemically abrupt interface between the film and STO substrate. On the other hand, the Ti-$L_{2,3}$ signal is seen to extend into the IL. Figure 4d shows background-subtracted Ti-$L$, V-$L$ and O-$K$ EELS profiles obtained layer-by-layer across the IL. Standard spectra obtained from single-crystalline VO$_2$(B) (V$^{4+}$) and V$_2$O$_3$ (V$^{3+}$) thin films are also included for comparison. The peak position of V-$L_{2,3}$ edges are seen to remain fixed indicating little to no change in the valence state of V in the IL and the VO$_2$(B) film.

The Ti-$L_{2,3}$ EELS fine structure obtained from the extra (Ti,V)-O layer on the STO substrate surface shows broadened $L_3$ and $L_2$ edges, as well as a shift of the $e_g$ peaks toward lower energy-loss (see Figure 4d). The observed electronic state and atomic structure of this extra layer are in good agreement with previous theoretical simulations[25] and STEM observations[26] of the c(4 × 2) reconstructed STO(001) surface composed of a double-layer TiO$_2$. The topmost layer of this reconstructed surface was predicted to contain clustered quartets of edge-sharing square-pyramidal TiO$_5$. It is probably that the extra (Ti,V)-O layer on TiO$_2$-terminated STO can introduce edge sharing oxygen containing units, which is more consistent with the VO$_2$(B) structure. To our knowledge, the formation of such an interface bi-layer is not ready to be rationalized by any the existing growth models that involve either phase transition[27-29] or phase separation[30-32] at film/substrate interfaces to accommodate inter-phase structural discontinuities.

The observed results reveal unambiguously, at the initial growth stage, the formation of VO$_2$(B)/STO heterostructure involves a structural reconstruction process at the substrate surface to facilitate the symmetry transition between the two distinct component structures, followed by the epitaxial growth of VO$_2$(B) nanodomains. The VO$_2$(B) nanodomains forms a fully coherent interface with the STO substrate and are subject to considerable lattice strain. Once the strain energy in the VO$_2$(B) nanodomains exceeds some critical level, misfit dislocations are introduced and the VO$_2$(B) film then continues to grow in a fully relaxed state. The much larger domain size in the relatively strain-free film is an expected result of increased adatom mobility on the relaxed surface. Formation of the interfacial VO$_2$(B) nanodomains indicates a nanoscale self-templating process that enables the epitaxy of strain-free VO$_2$(B) film on STO substrate. The results not only enable novel insights into atomic mechanism of complex heterostructure interface at an atomic scale, but also shed light on the epitaxial design of two materials with large symmetry and lattice mismatch.

**Methods**

*Epitaxial synthesis.* VO$_2$(B) epitaxial films were deposited on (001) SrTiO$_3$ substrates by pulsed laser epitaxy. A sintered ceramic VO$_2$ target was ablated with a KrF excimer laser ($\lambda$ = 248 nm) at a repetition rate of 5 Hz and laser fluence of 1 Jcm$^{-2}$. The optimized substrate temperature and oxygen pressure to grow high quality thin films were 500º C and 20 mTorr, respectively, and the samples were *in*-situ post-annealed in 1 atm of O$_2$ for 1 hour at the growth temperature to ensure the oxygen stoichiometry. Detailed information on the synthesis of single-crystalline VO$_2$(B) (V$^{4+}$) and V$_2$O$_3$ (V$^{3+}$) thin films utilized for EELS analysis can be found elsewhere.[14]

*Scanning Transmission Electron Microscopy (STEM).* Cross-sectional specimens oriented along the [100]STO direction for STEM analysis were prepared using ion milling after mechanical thinning and precision polishing (using water-free abrasive). High-angle annular dark-field (HAADF) and low-angle annular dark-field (LAADF) imaging and electron-energy loss spectroscopy (EELS) analysis were carried out in Nion UltraSTEM200 operated at 200 keV. The microscope is equipped with a cold field-emission gun and a corrector of third- and fifth-order aberrations for sub-angstrom resolution. Inner/outer detector angles of 78/240 mrad and 30/63 mrad were used for HAADF and LAADF imaging, respectively. The convergence semi-angle for the electron probe was set to 30 mrad.

**Figure Captions**

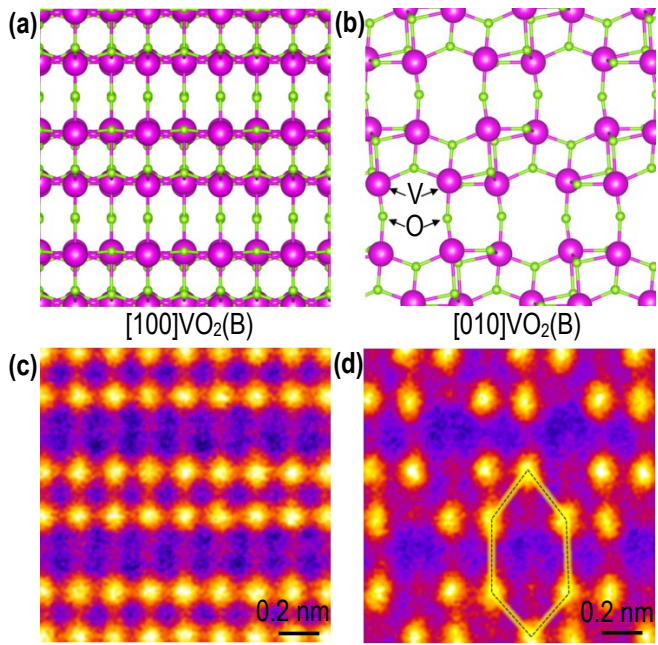

**Figure 1**. Atomic structure of $VO_2(B)$. a,b) Schematics and c,d) corresponding cross-sectional HAADF images of $VO_2(B)$ seen along a,c) the [100] and b,d) [010] directions. The hexagon in d) indicates the large open channel in $VO_2(B)$ useful for ionic conduction.

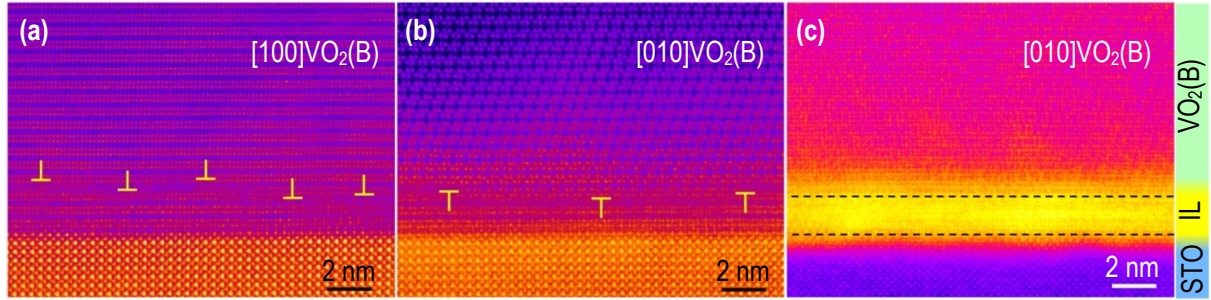

**Figure 2**. Microstructure of the VO$_2$(B)/STO interface. HAADF images show two growth twins orthogonally oriented along a) the [100]$_{VO2(B)}$ and b) [010]$_{VO2(B)}$ directions with respect to the [100]$_{STO}$ direction. c) LAADF image taken from the image in b), showing an extra intensity from the IL associated with increased electron beam dechanneling and, thus, scattering of electrons due to increased atomic disorder.

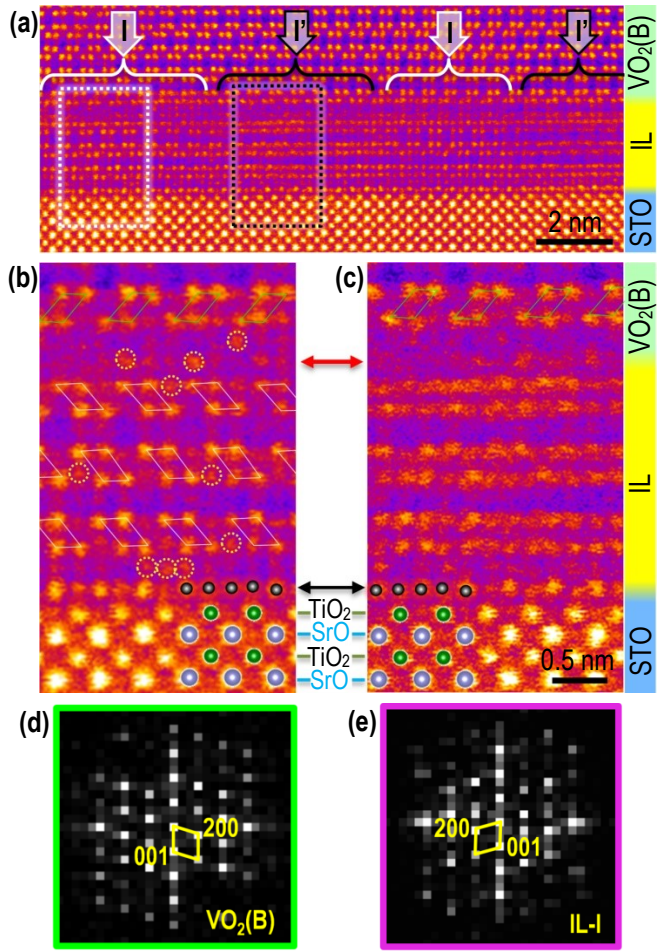

**Figure 3.** High-resolution observation of IL. a) HAADF image showing the IL consists of nanodomains, e.g. I and I'. b,c) Magnified HAADF images taken from nanodomains I and I' marked by the dashed rectangles in a) showing in greater detail the local atom arrangements. d,e) FFT electron diffraction patterns obtained the $VO_2(B)$ film and the IL (nanodomain-I), respectively. The red and black arrows between b) and c) indicate extra atom planes formed at the upper and lower sides of IL, respectively.

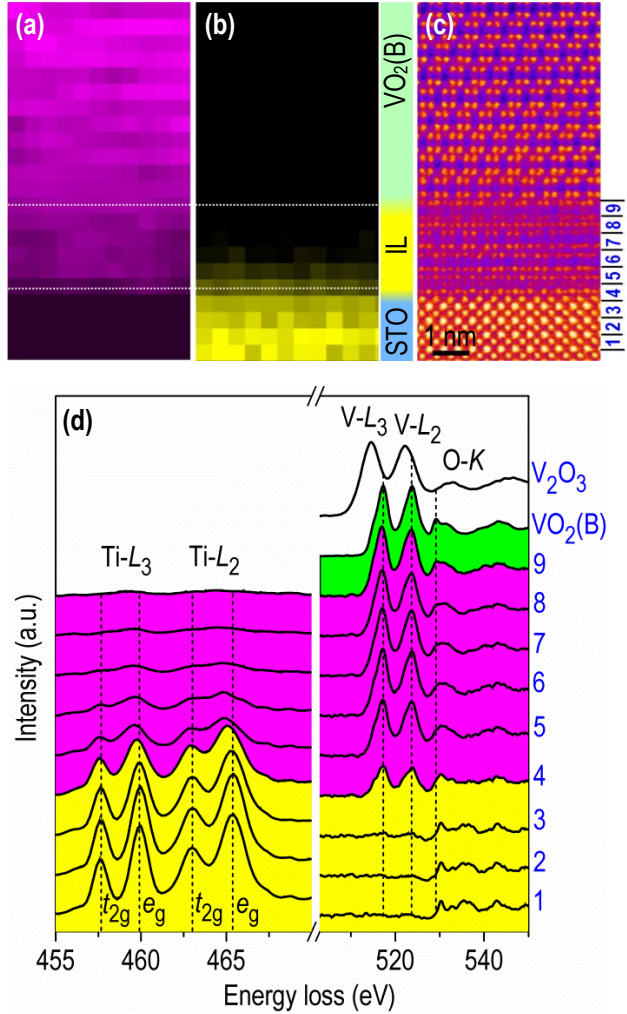

**Figure 4**. Layer-by-layer EELS analysis. Elemental maps for a) V-$L_{2,3}$ and b) Ti-$L_{2,3}$ signals obtained from the interface region shown in c). d) Back-ground subtracted Ti-$L_{2,3}$, V-$L_{2,3}$, and O-$K$ spectra obtained across the interface. The EELS spectra numbered 1 through 9 are obtained from the local atomic planes indicated in c). EELS profile intensity is normalized using the O-$K$ edge beyond the ionization transitions to discrete states.

**References**

1   Hwang, H. Y. *et al.* Emergent phenomena at oxide interfaces. *Nature materials* **11**, 103-113 (2012).

2   Zubko, P., Gariglio, S., Gabay, M., Ghosez, P. & Triscone, J.-M. Interface physics in complex oxide heterostructures. *Annu. Rev. Condens. Matter Phys.* **2**, 141-165 (2011).

3   Jeen, H. *et al.* Reversible redox reactions in an epitaxially stabilized SrCoOx oxygen sponge. *Nature materials* **12**, 1057-1063 (2013).

4   Gorbenko, O. Y., Samoilenkov, S., Graboy, I. & Kaul, A. Epitaxial stabilization of oxides in thin films. *Chemistry of materials* **14**, 4026-4043 (2002).

5   Morin, F. Oxides which show a metal-to-insulator transition at the Neel temperature. *Physical Review Letters* **3**, 34 (1959).

6   O'regan, B. & Grfitzeli, M. A low-cost, high-efficiency solar cell based on dye-sensitized. *nature* **353**, 737-740 (1991).

7   Arico, A. S., Bruce, P., Scrosati, B., Tarascon, J.-M. & Van Schalkwijk, W. Nanostructured materials for advanced energy conversion and storage devices. *Nature materials* **4**, 366-377 (2005).

8   Khan, S. U., Al-Shahry, M. & Ingler, W. B. Efficient photochemical water splitting by a chemically modified n-TiO2. *science* **297**, 2243-2245 (2002).

9   Yang, Z., Ko, C. & Ramanathan, S. Oxide electronics utilizing ultrafast metal-insulator transitions. *Annual Review of Materials Research* **41**, 337-367 (2011).

10  Park, J. H. *et al.* Measurement of a solid-state triple point at the metal-insulator transition in VO2. *nature* **500**, 431-434 (2013).

11   Morrison, V. R. *et al.* A photoinduced metal-like phase of monoclinic VO2 revealed by ultrafast electron diffraction. *science* **346**, 445-448 (2014).

12   Jeong, J. *et al.* Suppression of metal-insulator transition in VO2 by electric field–induced oxygen vacancy formation. *science* **339**, 1402-1405 (2013).

13   Luttrell, T. *et al.* Why is anatase a better photocatalyst than rutile?-Model studies on epitaxial TiO2 films. *Scientific reports* **4** (2014).

14   Lee, S., Meyer, T. L., Park, S., Egami, T. & Lee, H. N. Growth control of the oxidation state in vanadium oxide thin films. *Applied Physics Letters* **105**, 223515 (2014).

15   Chen, A. *et al.* Textured metastable VO2 (B) thin films on SrTiO3 substrates with significantly enhanced conductivity. *Applied Physics Letters* **104**, 071909 (2014).

16   Srivastava, A. *et al.* Selective growth of single phase VO2 (A, B, and M) polymorph thin films. *APL materials* **3**, 026101 (2015).

17   Lee, S., Ivanov, I. N., Keum, J. K. & Lee, H. N. Epitaxial stabilization and phase instability of VO2 polymorphs. *Scientific reports* **6** (2016).

18   Pokrovskii, B. & Khachaturyan, A. The concentration wave approach to the pairwise interaction model for predicting the crystal structures of ceramics, I. *Journal of Solid State Chemistry* **61**, 137-153 (1986).

19   Li, W., Dahn, J. R. & Wainwright, D. S. Rechargeable lithium batteries with aqueous electrolytes. *Science-AAAS-Weekly Paper Edition-including Guide to Scientific Information* **264**, 1115-1117 (1994).

20   Mai, L. *et al.* Nanoscroll Buffered Hybrid Nanostructural VO2 (B) Cathodes for High-Rate and Long-Life Lithium Storage. *Advanced materials* **25**, 2969-2973 (2013).

21  Niu, C. *et al.* VO2 nanowires assembled into hollow microspheres for high-rate and long-life lithium batteries. *Nano letters* **14**, 2873-2878 (2014).

22  Cowley, J. & Huang, Y. De-channelling contrast in annular dark-field STEM. *Ultramicroscopy* **40**, 171-180 (1992).

23  Hillyard, S. & Silcox, J. Detector geometry, thermal diffuse scattering and strain effects in ADF STEM imaging. *Ultramicroscopy* **58**, 6-17 (1995).

24  Pennycook, S. J. & Nellist, P. D. *Scanning transmission electron microscopy: imaging and analysis*. (Springer Science & Business Media, 2011).

25  Erdman, N. *et al.* Surface structures of SrTiO3 (001): A TiO2-rich reconstruction with a c(4× 2) unit cell. *Journal of the American Chemical Society* **125**, 10050-10056 (2003).

26  Zhu, G.-z., Radtke, G. & Botton, G. A. Bonding and structure of a reconstructed (001) surface of SrTiO3 from TEM. *nature* **490**, 384-387 (2012).

27  Lazarides, N., Paltoglou, V., Maniadis, P., Tsironis, G. & Panagopoulos, C. Strain-induced interface reconstruction in epitaxial heterostructures. *Physical Review B* **84**, 245428 (2011).

28  Zhou, H., Chisholm, M. F., Yang, T.-H., Pennycook, S. J. & Narayan, J. Role of interfacial transition layers in VO2/Al2O3 heterostructures. *Journal of Applied Physics* **110**, 073515 (2011).

29  Bayati, M. *et al.* Domain epitaxy in TiO2/α-Al2O3 thin film heterostructures with Ti2O3 transient layer. *Applied Physics Letters* **100**, 251606 (2012).

30  Lazarov, V. K., Chambers, S. A. & Gajdardziska-Josifovska, M. Polar oxide interface stabilization by formation of metallic nanocrystals. *Physical Review Letters* **90**, 216108 (2003).

31  Turner, S. *et al.* Structural phase transition and spontaneous interface reconstruction in La 2/3 Ca 1/3 MnO 3/BaTiO 3 superlattices. *Physical Review B* **87**, 035418 (2013).

32  Gao, X. *et al.* Structural Distortion and Compositional Gradients Adjacent to Epitaxial LiMn2O4 Thin Film Interfaces. *Advanced Materials Interfaces* **1** (2014).

**Acknowledgements**

This work was supported by the U.S. Department of Energy, Office of Science, Basic Energy Sciences, Materials Sciences and Engineering Division. We would like to thank Qian He and Erjia Guo for helpful discussions.

**Author Contributions**

The manuscript was written through contributions of all authors. X. G. conceived and designed the experiments under supervision of H.N.L. S.L. fabricated the film samples. All authors have given approval to the final version of the manuscript.

**Additional Information**

**Competing financial interests**: The authors declare no competing financial interests.